%% file: main.tex
\documentclass[runningheads,11pt]{llncs}

\input{macros}

\begin{document}

\title{Realizability Checking of Contracts with Kind 2 \\[1.5ex]
(Draft)
}

\author{
  Daniel Larraz\inst{1}
  \and
  Cesare Tinelli\inst{1}
}
\authorrunning{D. Larraz et al.}

\institute{
Department of Computer Science, The University of Iowa. USA
}

\maketitle

\begin{abstract}
We present a new feature of the open-source model checker Kind 2 which
checks whether a component contract is realizable; i.e.,
it is possible to construct a component such that for any input allowed by
the contract assumptions, there is some output value that the component
can produce that satisfies the contract guarantees.
When the contract is proven unrealizable, it provides a deadlocking computation
and a set of conflicting guarantees.
This new feature can be used to detect flaws in component specifications and
to ensure the correctness of Kind 2's compositional proof arguments.

\end{abstract}

\section{Introduction}
\label{sec:introduction}

Contract-based software development has long been a leading methodology for
the construction of component-based reactive systems,
embedded systems in particular.
Contracts provide a mechanism for capturing the information needed to
specify and reason about component-level properties at a desired level of
abstraction. In this paradigm, each component is associated with a contract
specifying its input-output behavior in terms of guarantees provided by
the component when its environment satisfies certain given assumptions.
Contracts are an effective way to establish boundaries between components and
can be used efficiently to prove global properties about a system prior to
its construction.
Such proofs are built upon the premise that each leaf-level component contract
in the system hierarchy is realizable; i.e., it is possible to construct
a component such that for any input allowed by the contract assumptions,
there is some output value that the component can produce that satisfies
the contract guarantees.
However, without engineering support it is all too easy to write
leaf-level components that cannot be realized.

This report describes a new feature of the open-source model checker
\kind~\cite{Kind2CAV16} which allows users to verify the realizability 
of contracts.
\kind is an SMT-based model checker for safety properties of finite- and 
infinite-state synchronous reactive systems. It takes as input models written 
in an extension of the Lustre language~\cite{Lustre92} that allows the specification of 
assume-guarantee-style contracts for system components. 
\kind's contract language~\cite{CoCoSpec16} is expressive enough to allow one 
to represent any (LTL) regular safety property by recasting it 
in terms of invariant properties.
One of \kind's distinguishing features is its support for modular
and compositional analysis of hierarchical and multi-component systems.
\kind traverses the subsystem hierarchy bottom-up, analyzing each system component,
and performing fine-grained abstraction and refinement of the sub-components.

The behavior of each component can be specified by providing either
a set of equations that define the component's output
in terms of its input and internal state (a \emph{low-level} specification), 
or an assume-guarantee contract (a \emph{high-level} specification), or both.
The syntax restrictions and semantics of the Lustre language ensure
that every low-level specification of a component is \emph{executable} 
in the sense that for each possible input for the component and 
each internal state there is a unique output and next state 
for the component to move to.
When both specifications are provided, the low-level specification
is expected to be a refinement of the high-level one.
\kind checks this by verifying that every execution that satisfies 
the former also satisfies the latter.
Informally, we say that the set of equations \emph{satisfy} the contract. 
However, in compositional reasoning, when only a contract is provided 
for a subcomponent,
\kind assumes the existence of a component satisfying the contract 
when checking the satisfaction of the top-level component requirements, 
which may lead to bogus compositional proof arguments 
when the subcomponent's contract is unrealizable.

\begin{figure}[t]
  \begin{lustreTinyMath}[1em]
type digit_range = subrange [0,9] of int; |\label{display:line:subrange}|
const MAX_TIME = 60 * 9 + 59;
    
node imported Display_Control( |\label{display:line:header}|
  cancel: bool; incr: bool; decr: bool; baking: bool |\label{display:line:inputs}|
)
returns (
  left_digit: digit_range; middle_digit: digit_range; right_digit: digit_range; |\label{display:line:digits}|
  minutes_to_cook: int |\label{display:line:minutes}|
);
\end{lustreTinyMath}
  \caption{Display\_Control component}
  \label{fig:displayControl}
\end{figure}

\begin{figure}[t]
  \begin{lustreTinyMath}[1em]
(*@contract |\label{display:line:contract:begin}|
  -- The following three definitions are based on the fact that
  -- minutes_to_cook shall match the total time of the displayed digits
  guarantee "G1: The left-most digit corresponds to hours" |\label{display:line:G1:begin}|
    left_digit = (minutes_to_cook div 60);
  guarantee "G2: The middle digit corresponds to tens of minutes"
    middle_digit = (minutes_to_cook mod 60) div 10;
  guarantee "G3: The right digit corresponds to minutes"
    right_digit = (minutes_to_cook mod 10); |\label{display:line:G3:end}|
    
  var any_button_pressed: bool = incr or decr or cancel; -- auxiliary var
    
  guarantee "G4: minutes_to_cook shall be initially zero"
    minutes_to_cook = 0 -> true;
  guarantee "G5: If the cancel button is pressed, minutes_to_cook shall be zero"
    cancel => minutes_to_cook = 0;
  guarantee "G6: When baking, minutes_to_cook shall remain the same or decrease"
    true -> baking => minutes_to_cook <= pre minutes_to_cook;
  guarantee "G7: When not baking, if not button is pressed, minutes_to_cook shall not change"
    true -> (not baking and not any_button_pressed) => minutes_to_cook = pre minutes_to_cook;
  guarantee "G8: When not baking, if incr is pressed, minutes_to_cook shall
             increase by one if it was less than MAX_TIME or be zero otherwise"
    true -> (not baking and incr) =>
      (minutes_to_cook = if pre minutes_to_cook < MAX_TIME then pre minutes_to_cook + 1 else 0);
  guarantee "G9: When not baking, if decr is pressed but not incr, minutes_to_cook shall
             decrease by one if it was greater than 0 or be MAX_TIME otherwise"
    true -> (not baking and not incr and decr) => 
      (minutes_to_cook = if pre minutes_to_cook > 0 then pre minutes_to_cook - 1 else MAX_TIME);
*) |\label{display:line:contract:end}|
\end{lustreTinyMath}
  \caption{\kind's contract for the Display\_Control component}
  \label{fig:displayControlContract}
\end{figure}

\begin{example}
We will use a simple model to illustrate the concepts and the functionality of \kind introduced in
this report. Suppose we want to design a component that controls the display of an oven.
The oven has a panel with three buttons: a \emph{cancel} button, an \emph{increase} button, and
a \emph{decrease} button.
The component that controls the display reads the three button inputs from the panel and
the current mode of the oven (baking or not baking), and
it sets three digit displays showing the current number of minutes to cook accordingly.
The left-most digit corresponds to hours, the middle digit is tens of minutes, and
the right digit is minutes.

Our model for the component is described in \kind's input language in
Figure~\ref{fig:displayControl}, which defines (starting at line~\ref{display:line:header}) the component's interface,
and Figure~\ref{fig:displayControlContract}, which contains its contract.
The interface includes three inputs, \code{cancel}, \code{incr}, and \code{decr},
one for each button input, and an additional input, \code{baking},
to indicate whether the oven is in baking mode or not
(line~\ref{display:line:inputs}).
The component has one output for each of the three digit displays
(line~\ref{display:line:digits})
whose values range between $0$ and $9$ (line~\ref{display:line:subrange}).
In addition, the component has an additional output that reports
the total time of the displayed digits (line~\ref{display:line:minutes}).
Following a model-based design, we model an abstraction of
\code{Display\_Control} component instead of specifying a complete set of
equations that fully determine the behavior of the whole component.
\kind allows the user to specify contracts for individual nodes,
either as special Lustre comments added directly inside the node declaration,
or as the instantiation of an external stand-alone contract that can be imported in
the body of other contracts.
The contract of \code{Display\_Control}, included directly in the node
(lines~\ref{display:line:contract:begin}-\ref{display:line:contract:end} 
of Figure~\ref{fig:displayControlContract}),
specifies the relationship between the value of \code{minutes\_to\_cook} and
the three displayed digits (guarantees \code{G1}-\code{G3})
as well as the value of \code{minutes\_to\_cook} in reaction to different
situations (guarantees \code{G4}-\code{G9}).
For instance, guarantee \code{G8} specifies that 
\code{minutes\_to\_cook} shall increase by one, when the oven is not cooking,
if the \code{incr} button is pressed and the previous total time to cook is
less than \code{MAX\_TIME}. Moreover, when the oven is not baking,
\code{minutes\_to\_cook} shall be zero if the \code{incr} button is pressed
but the previous total time to cook is not less than \code{MAX\_TIME}.

This specification is detailed enough to prove some interesting properties about
the component. In order to do that, we can wrap up an instance of
the \code{Display\_Control} component in an \emph{observer} component,
specify the properties we want to check as guarantees in the contract of
the observer component, and ask \kind to check the satisfaction of the contract.
For example, the contract of \code{Display\_Control\_Observer}
in Figure~\ref{fig:observer} specifies three properties
(\code{P1}-\code{P3}) about the behavior of \code{Display\_Control}.
\kind is able to prove the satisfaction of the three properties, however
\kind reasoning is oblivious to the fact that there are two guarantees
in the contract of \code{Display\_Control} that make the specification
unrealizable.
The new feature of \kind for checking the realizability of contracts is able
to detect that.

In addition, \kind provides a deadlocking computation and
a set of conflicting guarantees 
to help the designer identify the source of the problem.
In particular, in this case \kind returns a deadlocking computation
where initially all inputs are $\false$ and
\code{minutes\_to\_cook} is $0$, and, at the next step,
both \code{cancel} and \code{decr} are $\true$, the rest of inputs are $\false$,
and \code{minutes\_to\_cook} is $599$.
It also reports that, for the provided deadlocking computation,
guarantees \code{G5} and \code{G9}
form a minimal set of conflicting guarantees since, 
when both \code{cancel} and \code{decr} are $\true$ simultaneously,
\code{minutes\_to\_cook} cannot always be zero (\code{G5}) and be decreased 
by one (\code{G9}) at the same time.
One way to fix this issue is to update guarantee \code{G9} to strengthen
the premises of the implication with the requirement that
the \code{cancel} button is not pressed.
If the realizability of the contract is analyzed again after this update,
an analogous conflict is found between \code{G5} and \code{G8}.
Figure~\ref{fig:fixedGuarantees} shows guarantees \code{G8} and \code{G9}
after both contracts have been updated.
The contract with the revised guarantees is proven realizable by \kind.
Moreover, \kind is still able to prove properties \code{P1}-\code{P3}.
\end{example}

\begin{figure}[t]
  \begin{lustreTinyMath}[3em]
node Since ( X, Y : bool ) returns ( Z : bool ) ;
let
  Z =  X or (Y and (false -> pre Z)) ;
tel
    
node Once(x : bool) returns (Y : bool);
let
  Y = x or (false -> pre Y);
tel

node Display_Control_Observer(
  cancel: bool; incr: bool; decr: bool; baking: bool
)
returns(
  left_digit: digit_range; middle_digit: digit_range; right_digit: digit_range;
  minutes_to_cook: int
);
(*@contract
  var ctime: int = (left_digit*60) + (middle_digit*10) + right_digit;

  guarantee "P1: minutes_to_cook shall match the total time of the displayed digits"
    minutes_to_cook = ctime;
  guarantee "P2: minutes_to_cook shall not exceed the MAX_TIME value"
    minutes_to_cook <= MAX_TIME;
  guarantee "P3: If ctime is equal one, and since ctime was greater or equal than two, incr has not been pressed and the oven has not been baking, then decr must have been pressed once in the past"
    Since(ctime>=2, not incr and not baking) and ctime=1 => Once(decr);
*)
let
  left_digit, middle_digit, right_digit, minutes_to_cook =
    Display_Control(cancel, incr, decr, baking);
tel
\end{lustreTinyMath}
  \caption{\kind's contract for the Observer component}
  \label{fig:observer}
\end{figure}

\begin{figure}[h]
  \begin{lustreTinyMath}[3em]
guarantee "G8: When not baking, if incr is pressed but not cancel, minutes_to_cook shall increase by one if it was less than MAX or be zero otherwise"
  true -> 
    (not baking and not cancel and incr) =>
    (minutes_to_cook = if pre minutes_to_cook < MAX then pre minutes_to_cook + 1 else 0);

guarantee "G9: When not baking, if decr is pressed but not cancel nor incr, minutes_to_cook shall decrease by one if it was greater than 0 or be MAX otherwise"
  true ->
    (not baking and not cancel and not incr and decr) => 
    (minutes_to_cook = if pre minutes_to_cook > 0 then pre minutes_to_cook - 1 else MAX);
\end{lustreTinyMath}
  \caption{Guarantees G8 and G9 updated to make Display\_Control's contract realizable}
  \label{fig:fixedGuarantees}
\end{figure}

\section{Preliminaries}
\label{sec:preliminaries}

Lustre is a synchronous dataflow language that allows one to define system components 
as \define{nodes},
each of which maps a continuous stream of inputs 
(of various basic types, such as Booleans, integers, and reals)
to continuous streams of outputs based on both current input values and 
previous input and output values.
Bigger components can be built by parallel composition of smaller ones, achieved 
syntactically with \define{node applications}. 
Operationally, a node has a cyclic behavior: at each tick $t$ of a global clock
(or a local clock it is explicitly associated with)
it reads the value of each input stream at position or \define{time} $t$, and
instantaneously computes and returns the value of each output stream at time $t$. 

Formally, a stream of values of type $\tau$ is a function the natural numbers
(modeling the global clock ticks) to $\tau$. 
The behavior of a Lustre node is then specified declaratively 
by a set of stream constraints
of the form $x = s$, where $x$ is a variable denoting an output or a locally
defined stream and $s$ is a stream term over input, output, and local variables.
Most stream operators are point-wise liftings of the usual operators over stream
values.
For example, if $x$ and $y$ are two integer streams, 
the term $x + y$ is the stream corresponding the function 
$\lambda t.\, x(t) + y(t)$ over time $t$; 
an integer constant $c$, denotes the constant function $\lambda t.\, c$. 
Two important additional operators are a unary right-shift operator \code{pre}, used
to specify stateful computations, and a binary initialization operator \code{->},
used to specify initial state values. 
At time $t=0$, the value $(\code{pre}\ x)(t)$ is undefined; 
for each time $t > 0$, it is $x(t-1)$.
In contrast, the value $(x \ \code{->}\ y)(t)$ equals $x(t)$ for $t = 0$
and $y(t)$ for $t > 0$.
Syntactic restrictions guarantee that all streams in a node are inductively well defined. 
In Kind 2's extension of Lustre, nodes can be given assume-guarantee contracts,
enabling the compositional analysis of Lustre models.
Contracts specify assumptions as Boolean terms over current values of input streams and 
previous values of input and output streams, and 
guarantees as Boolean terms over current and previous values of input and output streams.

After various transformations and slicing, \kind
encodes Lustre nodes internally as (state) transition systems $\sys=\sysTuple$ 
where $\vS$ is a vector of typed state variables,
$\vI$ is a vector of typed input variables,
$\init$ is the initial state predicate (over the variables in $\vS$ and $\vI$), 
and $\trans$ is a two-state transition predicate
(over the variables in $\vS$, $\nxtI$ and $\nxtS$, 
with $\nxtI$ and $\nxtS$ being a renamed version of $\vS$ and $\vI$,
respectively).
System outputs are represented by selected elements of $\vS$ 
which we do not distinguish from internal state variables for simplicity.
We will use $\pair{\init}{\trans}$ to refer to transition system $\sys$
when the vectors of state and input variables, $\vS$ and $\vI$,
are clear from the context or not important.
A \define{contract} for $\sys$ is a pair $\agContract = \agTuple$
of an \define{assumption} transition system $\ass=\assTuple$,
where $\vI$ and $\vS$ act, respectively, as the state variables and
the input variables of the environment of $\sys$,\footnote{For simplicity, 
but without loss of generality, we assume that any reference to
a previous value of an input variable in $\assT$ is made through 
a state variable in $\vS$ storing that value, and thus, 
$\assT$ is defined only over $\vS$ and $\nxtI$.}
and
a \define{guarantee} transition system $\gua=\guaTuple$ with
the same state and input variables as transition system $\sys$.
The predicates $\assI$ and $\assT$ specify, respectively, which
inputs are {\em valid} initially and for a given system state.
The predicate $\guaI$ specifies which states the system may start in
when the initial inputs satisfy $\assI$.
The predicate $\guaT$ specifies for a given state and 
inputs what states the system may transition to
when the inputs satisfy $\assT$.
Given a contract $\agContract = \agTuple$,
we will assume that $\guaI$ and $\guaT$ have the structure of a top-level conjunction, 
that is, $\guaT[\vS, \nxtI, \nxtS] =
{\guaT}_1[\vS, \nxtI, \nxtS] \land \cdots \land {\guaT}_n[\vS, \nxtI, \nxtS]$
for some $n\geq 1$. Notice that Kind 2's assume-guarantee contracts 
follow naturally this kind of conjunctive structure since 
they are specified as conjunction of assumptions and a conjunction
of guarantees. 
By a slight abuse of notation, we will identify $\guaI$ ($\guaT$),
with the set $\{{\guaI}_1, \ldots, {\guaI}_m\}$ 
($\{{\guaT}_1, \ldots, {\guaT}_n\}$) of its top-level conjuncts.

Now we will introduce some definitions and results required to describe
the new functionality of \kind.
Given a vector of typed variables $\vX$, a {\em valuation} $\nu$ over $\vX$
is a type-consistent assignment of values to all the variables in $\vX$.
For a valuation $\nu$ over a vector $\vX$, 
we denote by $\nu'$ the valuation such that 
$\nu'(x') = \nu(x)$ for all variables $x$ in $\vX$; 
for a vector $\vY$ consisting of variables from $\vX$,
we denote by $\nu[\vY]$ the valuation over $\vY$ obtained by restricting $\nu$ 
to the variables in $\vY$.
Given a transition system $\sys=\sysTuple$, a \define{state} of $\sys$ is
a valuation over $\vS$ and an \define{input} of $\sys$ is a valuation over $\vI$.
A \define{trace} $\sigma$ is a sequence of valuations over $\vS$ and $\vI$.
A \define{computation path $\pi$ of $\sys$ of length $k\geq 0$} is 
a finite sequence of valuations $\pi=\pi_0,\ldots,\pi_k$
over $\vS$ and $\vI$ such that $\pi_0$ satisfies the predicate $\init[\vS,\vI]$,
and for every $0\leq i < k$ the valuations $\pi_i,\pi'_{i+1}$ satisfy
the predicate $\trans[\vS, \nxtI, \nxtS]$.
A \define{trace of $\sys$} is a trace $\sigma=\sigma_0,\sigma_1,\ldots$ such that
for every $k\geq 0$ the prefix $\sigma_0,\ldots,\sigma_k$ is
a computation path of $\sys$.
We will denote the set of all the traces of $\sys$ as $L(\sys)$.
Given two transition systems $\sys_1$ and $\sys_2$ with
the same vectors of state and inputs variables,
$\sys_2$ is a (trace-based) \define{refinement} of $\sys_1$ iff
$L(\sys_2)\subseteq L(\sys_1)$.

The following definitions are adapted from similar notions introduced
by Gacek et al.~\cite{Gacek15}.
Unlike the original paper, we explicitly formalize the fact that
predicate $\guaI$ may depend on input values, and
that assumptions may specify constraints over the initial input values
through the predicate $\assI$.
This allows for greater generality and flexibility
without significantly affecting the context or the proven results.
In the following, 
we fix for convenience a transition system $\sys=\sysTuple$.
In the definitions below, we will consider 
\begin{itemize}
\item assumptions $\ass$ of the form $\assTuple$ and
\item guarantees $\gua$ of the form $\guaTuple$
\end{itemize} 
for $\sys$.

\begin{mydef}
A computation path $\pi=\pi_0,\ldots,\pi_k$ of $\sys$
\define{satisfies an assumption $\ass$}
if $\pi_0$ satisfies the predicate $\assI[\vI]$, and
for every $0\leq i < k$ the valuations $\pi_i,\pi'_{i+1}$ satisfies
the predicates $\assTP$.
\end{mydef}

\begin{mydef}
A state $\vRS$ is \define{reachable (in $\sys$) under an assumption}
$\ass$ if there exists a computation path $\pi=\pi_0,\ldots,\pi_k$ of $\sys$
satisfying $\ass$ such that $\pi_k[\vS]=\vRS$.
Formally, the set $\reachableSA(\vS)$
of reachable states under an assumption $\ass$
 is defined inductively by the following equation:
\[\reachableSA(\vS)\triangleq(\exists \vI_0.\, \assI[\vI_0] \land \init[\vS]) \lor 
(\exists \vS_p,\nxtI.\, \reachableSA(\vS_p) \land \assT[\vS_p,\nxtI] \land \trans[\vS_p,\nxtI,\vS])\]
\end{mydef}

\begin{mydef}
The transition system $\sys$ \define{satisfies} a contract $\agContract = \agTuple$
when the following conditions hold:
\begin{enumerate}
  \item $\forall \vS,\vI.\, \assI[\vI] \land \init[\vS,\vI] \Rightarrow \guaI[\vS,\vI]$
  \item $\forall \vS,\nxtI,\nxtS.\, \reachableSA(\vS) \land \assTP \land
         \trans[\vS, \nxtI, \nxtS] \Rightarrow \guaT[\vS, \nxtI, \nxtS]$
\end{enumerate}
\end{mydef}

When $\sys$ does not satisfy a contract $\agContract = \agTuple$, 
there is a computation path $\pi=\pi_0,\ldots,\pi_k$ of $\sys$ 
satisfying assumption $\ass$ such that
either $k=0$ and $\pi_0$ does not satisfy $\guaI[\vS,\vI]$ or
$k>0$, $\pi_0,\ldots,\pi_{k-1}$ is a computation path of $\gua$
but $\pi_{k-1},\pi'_k$ does not satisfy $\guaT[\vS, \nxtI, \nxtS]$.
We call such computation path a \define{safety counterexample},
and any trace that has that computation path as a prefix
a \define{safety counter-trace}.

\begin{mydef}
The transition system $\sys$ is \define{input-enabled under an assumption}
$\ass$ when the following two conditions hold:
\begin{enumerate}
  \item $\forall \vI.\, \assI[\vI] \Rightarrow \exists \vS.\, \init[\vS,\vI]$
  \item $\forall \vS,\nxtI.\, \reachableSA(\vS) \land \assTP \Rightarrow
         \exists \nxtS.\, \trans[\vS, \nxtI, \nxtS]$
\end{enumerate}
\end{mydef}

\begin{mydef}
The transition system $\sys$ is a \define{realization} of a contract $\agContract=\agTuple$
if $\sys$ satisfies $\agContract$ and is input-enabled under assumption $\ass$.
\end{mydef}

\begin{mydef}
A contract is \define{realizable} if there exists a transition system
which is a realization of the contract.
\end{mydef}

When a contract $\agContract=\agTuple$ is unrealizable, 
we can try to build an \define{environment} transition system $\env=\envTuple$
such that $E$ is a realization of contract $\pair{\emptyTuple}{\ass}$,
thus $E$ is a refinement of $A$ which always keeps running,
and $L(\env) \subseteq \overline{L(\gua)}$.
We call $E$ a \define{counter-strategy}.
A user can examine a counter-strategy to try understand 
the reasons the contract is unrealizable and fix it accordingly.
However, as pointed out by K{\"{o}}nighofer et al~\cite{KonighoferHB13},
a counter-strategy may be very large and complex.
Hence, the user may prefer a single computation path $\pi=\pi_0,\ldots,\pi_k$ of
$G$ satisfying $A$ such that state $\pi_k[\vS]$ satisfies
\begin{equation} \label{eq:deadlock}
 \exists \nxtI. \assTP \land \forall \nxtS.\neg \guaT[\vS,\nxtI,\nxtS]
\end{equation}
We say that $\pi_k[\vS]$ is a \emph{deadlocked} state.

Since knowing concrete input values for the existentially quantified variables 
in (\ref{eq:deadlock}) is relevant to understand why $\guaT$ cannot be satisfied,
instead of giving the user the computation path above,
we return an extended version of it. Namely,
our algorithm generates computation path 
$\hat{\pi}=\pi_0,\ldots,\pi_k,\pi_{k+1}$ such that
$\pi_k[\vS]$ and $\pi'_{k+1}[\nxtI]$ satisfy
$\assTP \land \forall \nxtS. \neg \guaT[\vS,\nxtI,\nxtS]$.
When an initial state does not always exist,
the algorithm, instead, generates
a computation path $\rho=\rho_{0}$ such that $\rho_{0}[\vI]$
satisfies $\assI[\vI] \land \forall \vS. \neg \guaI[\vS,\vI]$.
We will call such a computation path a \define{deadlocking computation},
and any trace that has the computation path as a prefix
a \define{realizability counter-trace}.
Although $\pi'_{k+1}$ ($\rho_{0}$) may give arbitrary values to
$\nxtS$ ($\vS$), our algorithm computes it so that a minimal set $U$
of guarantee conjuncts are violated,
where $U\subseteq\guaI$ when the violation happens at the initial step, and 
$U\subseteq\guaT$ when it happens later.
We call such a subset a \define{set of conflicting guarantees} or,
simply, a \define{conflict}.

The realizability check presented in this paper is based on a notion
called \emph{viability} introduced by Gacek et al.~\cite{Gacek15}, 
which provides a characterization of contract realizability.

\begin{mydef}
A state $\vRS$ is \define{viable with respect to} a contract $\agContract$,
if $\guaT$ can keep responding to valid inputs forever,
starting from $\vRS$. 
Formally, the set of viable states with respect to $\agContract$ is defined
coinductively by the following equation:
\[\viableC(\vS) \triangleq \forall \nxtI.\, \assT[\vS,\nxtI] \Rightarrow
  \exists \nxtS.\, \guaT[\vS, \nxtI, \nxtS] \land \viableC(\nxtS)\]
\end{mydef}

\begin{theorem}
A contract $\agContract$ is realizable if and only if
$\forall \vI.\, \assI[\vI] \Rightarrow \exists \vS.\, \guaI[\vS,\vI] \land \viableC(\vS)$ holds.
\label{thm:realizable}
\end{theorem} 

\begin{proof} Follows from our definition of input-enabled transition system
and an analogous proof to the one provided for Theorem 1 in~\cite{Gacek15}.\end{proof}

\section{An Algorithm for Checking Realizability}

In this section we present the algorithm used by \kind for automatically checking
the realizability of a contract, and finding a deadlocking computation and a conflict 
when the contract is proven unrealizable.
It is an adaptation to \kind of a synthesis procedure
by Katis et al.~\cite{Katis18}.
The algorithm iteratively refines an over-approximation of the set of viable states,
expressed as a predicate $F$, until $F$ is determined to be a fixpoint by proving
the validity of the following formula:

\begin{equation}
  \forall \vS,\nxtI.\, (F[\vS] \land \assTP
\Rightarrow \exists \nxtS.\, \guaT[\vS,\vI,\nxtS] \land F[\nxtS])
\label{eq:fixpoint-check}
\end{equation}

After the greatest fixpoint is computed, the realizability of the contract can be established
by checking whether for all initial valid inputs there exists a state that satisfies
$\guaI$ and $F$. When that is the case, the contract is realizable. Otherwise,
the contract is unrealizable.

To decide the validity of $\forall\exists$-formulas, the main algorithm relies
on the \aeval procedure (described in Algorithm~\ref{alg:aeval}).
\aeval starts computing a \define{region of validity} for the input formula,
i.e., a formula $P[\vec{x}]$ such that 
$\forall \vec{x}.\, Q[\vec{x}] \land P[\vec{x}] \Rightarrow \exists \vec{y}.\,W[\vec{x},\vec{y}]$
is valid.
It achieves that by applying quantifier elimination to
the formula $\exists \vec{y}.\,Q[\vec{x}] \Rightarrow W[\vec{x},\vec{y}]$
which takes into account the context $Q[\vec{x}]$ (line~\ref{aeval:line:qe}).
Then, it checks whether the formula $P[\vec{x}]$ is valid by checking if its
negation is unsatisfiable. If it is, the original formula is valid.
Otherwise, the original formula is invalid.
In both cases, the algorithm conjoins the computed region of validity with
$Q[\vec{x}]$ 
and then, 
it returns the region together with a Boolean value indicating the validity result.

\begin{algorithm}
  \caption{\textsc{AE-VAL} ($\forall \vec{x}.\, Q[\vec{x}] \Rightarrow \exists \vec{y}.\,W[\vec{x},\vec{y}]$)}
  \begin{algorithmic}[1]
  
  
  
  
  
  
  
  
  
  
  
  
  
  
     \State $P[\vec{x}] \gets \proc{QE}{\exists \vec{y}.\,Q[\vec{x}] \Rightarrow W[\vec{x},\vec{y}]}$ \label{aeval:line:qe}
  
     \State $\proc{SmtAssert}{\neg P[\vec{x}]}$
  
     \State $sat \gets \proc{SmtCheckSat}{ }$ \label{aeval:line:checksat}
  
     \State \Return $\pair{\neg sat}{Q[\vec{x}] \land P[\vec{x} ]}$
  
  
  \end{algorithmic}
  \label{alg:aeval}
  \end{algorithm}

The realizability check procedure is described in
Algorithm~\ref{alg:RealizabilityCheck}. 
It begins by checking that
there exists a state satisfying $\guaI$ for all initial valid inputs 
(line~\ref{rc:line:aevalI1}).
When that is not the case, the contract is unrealizable and a deadlocking computation
is generated together with a set of conflicting guarantees 
(line~\ref{rc:line:counter-traceI}).
This check can be seen as an optimization for detecting unrealizable contracts
without having to compute $F$, but it also helps to handle separately
the generation of a deadlocking computation for the initial case and
the transition case (line~\ref{rc:line:ctraceT}).
Then, the algorithm checks whether the contract is trivially realizable because
there are no initial valid inputs (line~\ref{rc:line:assIunsat}).
If it is the case, the algorithm terminates declaring
the contract realizable (line~\ref{rc:line:trvRealizable}).
Otherwise, it initializes four variables
before entering the main loop (line~\ref{rc:line:initialization}):
$F$ represents the current candidate fixpoint,
$\mvar{fl}$ is a flag that indicates whether $F$ has been refined
at least once, and $R$ and $\hat{R}$ are used to store 
(after the first refinement) the regions of
validity over $\vS$ and $\nxtI$, and $\vS$, respectively, for which
there exists a next state satisfying $\guaT$.
Both $R$ and $\hat{R}$ are arbitrarily initialized to $\top$.

In each iteration, the algorithm proceeds as follows.
First, it checks whether greatest fixpoint has been reached
by checking the validity of Formula~\ref{eq:fixpoint-check}
(line~\ref{rc:line:aevalT}).
If the formula is invalid, \aeval provides a region of validity
$\validRegionP$ over $\vS$ and $\nxtI$. This formula
may contain constraints over the contract's inputs, 
so it cannot be used to refine $F$ directly.
To determine the specific region over $\vS$ for which there exists
an input that violates Formula~\ref{eq:fixpoint-check}, we can use
\aeval again to determine the validity of formula 
$\phi' \gets \forall \vS.\, (F[\vS] \Rightarrow
\exists \nxtI.\, \assTP \land \neg \validRegionP)$.
The invalidity of $\phi'$ indicates that there are still non-violating states 
(i.e., outside $ violatingRegion[\vS]$) which may lead to a fixpoint.
Thus, the algorithm removes the unsafe states from $F[\vS]$ in 
line~\ref{rc:line:refinement}, and iterates until a greatest fixpoint 
for $F[\vS]$ is reached.
If $\phi'$ is valid, then every state in $F[\vS]$ is unsafe, 
under a specific input that satisfies the contract assumptions 
(since $\neg \validRegionP)$ holds in this case), and
the specification is unrealizable.
In the next iteration, the algorithm will reach 
line~\ref{rc:line:unrealizable}.
In addition, when $\mvar{fl}$ is $\false$ (i.e. it is the first iteration), 
the algorithm records 
$\validRegionP$ and $\neg violatingRegion[\vS]$ which are
used to generate a deadlocking computation and a conflict 
if the contract is determined to be unrealizable.

If $\phi$ is valid, the algorithm checks whether
for all initial valid inputs there exists a state that satisfies 
$\guaI$ and $F$ (line~\ref{rc:line:aevalI2}). 
If so, the the contract is realizable and the algorithm returns 
the generated fixpoint (line~\ref{rc:line:realizable}).
Otherwise, the contract is unrealizable and the algorithm generates
a deadlocking computation and a set of conflicting guarantees.
Since the algorithm has already verified that it is always
possible to compute an initial state for any valid initial input
(line~\ref{rc:line:valid1}), any counter-trace must involve
one or more states. Moreover, $F$ must has been refined
at least once, and $R$ and $\hat{R}$ set to
the regions of validity over $\vS$ and $\nxtI$, and $\vS$,
respectively, for which there exists a next state
satisfying $\guaT$ (lines~\ref{rc:line:setR}-\ref{rc:line:setHR}).
To generate the deadlocking computation, we must find 
a computation path of $\gua$ satisfying $\ass$ that reaches
a state $\vRS$ from which
it is impossible to transition to a new state satisfying $\guaT$,
i.e. $\models \neg\hat{R}[\vRS]$.
To find such computation path, the algorithm relies on a \verify procedure
that receives a transition system $\sys=\pair{\init}{\trans}$
and a contract $\agContract$, and returns a pair
$\pair{r}{c}$ where $r$ indicates whether $\sys$
satisfies the contract $\agContract$ or not, and
$c$ is a safety counterexample when $\sys$ does not satisfy
$\agContract$. The algorithm use \verify to check whether
transition system $\pair{\guaI}{\guaT}$ satisfy contract
$\pair{\ass}{\hat{R}}$.
Because the contract is unrealizable, it is ensured that
the call to \verify in line~\ref{rc:line:verify} always 
determines that $\sys$ does not satisfy $\agContract$ and
it returns a counterexample satisfying the properties stated above.

\begin{algorithm}
\caption{\textsc{RealizabilityCheck} ($A=\pair{\assI}{\assT}$, $G=\pair{\guaI}{\guaT}$)}
\begin{algorithmic}[1]

\State $\varphi \gets \forall \vI.\, \assI[\vI] \Rightarrow
        \exists \vS. \guaI[\vS,\vI]$

\State $\pair{valid}{validRegion[\vI]} \gets \Call{AE-VAL}{\varphi}$ \label{rc:line:aevalI1}

\If{$\neg valid$} \label{rc:line:valid1}

  \State $cex, C \gets \Call{GetDeadlockingCompAndConflict}{validRegion[\vI], \proc{EmptyList}{}, \assI, \guaI}$
  \label{rc:line:counter-traceI}

  \State \Return $\pair{ \textsc{\scriptsize{UNREALIZABLE}} }{ \pair{cex}{C} }$

\EndIf

\If{$\proc{IsUNSAT}{\assI[\vI]}$} \label{rc:line:assIunsat}

  \State \Return $\pair{ \textsc{\scriptsize{REALIZABLE}} }{ \top }$ \label{rc:line:trvRealizable}

\EndIf

\State $F[\vS] \gets \top$; $R[\vS,\nxtI] \gets \top$; $\hat{R}[\vS] \gets \top$;
       $\mvar{fl} \gets \false$ \label{rc:line:initialization}

\While{$\true$}

  \State $\phi \gets \forall \vS,\vI.\, (F[\vS] \land \assTP
           \Rightarrow \exists \nxtS.\, \guaT[\vS,\vI,\nxtS] \land F[\nxtS])$

  \State $\pair{ valid }{ \validRegionP } \gets \Call{AE-VAL}{\phi}$ \label{rc:line:aevalT}

  \If{$valid$}

    \State $\phi' \gets \forall \vI.\, \assI[\vI] \Rightarrow
             \exists \vS. \guaI[\vS,\vI] \land F[\vS]$

    \State $\pair{ valid' }{ \_ } \gets \Call{AE-VAL}{\phi'}$ \label{rc:line:aevalI2}

    \If{$valid'$}

      \State \Return $\pair{ \textsc{\scriptsize{REALIZABLE}} }{ F[\vS] }$ \label{rc:line:realizable}

    \Else

      \State $\_, cex \gets \proc{Verify}{\pair{\guaI}{\guaT}, \pair{\ass}{\hat{R}} \rangle}$ \label{rc:line:verify}

      \State $cex', C \gets \Call{GetDeadlockingCompAndConflict}{R, cex, \assT, \guaT}$ \label{rc:line:ctraceT}

      \State \Return $\pair{ \textsc{\scriptsize{UNREALIZABLE}} }{ \pair{cex'}{C} }$ \label{rc:line:unrealizable}

    \EndIf

  \Else
    \State $\phi' \gets \forall \vS.\, (F[\vS] \Rightarrow
            \exists \nxtI.\, \assTP \land \neg \validRegionP)$

    \State $\pair{ \_ }{ violatingRegion[\vS] } \gets \Call{AE-VAL}{\phi'}$

    \State $F[\vS] \gets F[\vS] \land \neg violatingRegion[\vS]$ \label{rc:line:refinement}

    \If{$\neg\mvar{fl}$}
      \State $R \gets \validRegionP$ \label{rc:line:setR}
      \State $\hat{R} \gets \neg violatingRegion[\vS]$ \label{rc:line:setHR}
      \State $\mvar{fl} \gets \false$
    \EndIf

  \EndIf

\EndWhile

\end{algorithmic}
\label{alg:RealizabilityCheck}
\end{algorithm}

To help the user to understand why a contract is unrealizable,
Algorithm~\ref{alg:GetDeadlockingCompAndConflictStar} computes
a set of conflicting guarantees, and a valuation for the inputs
and the state variables such that it satisfies as many guarantees
as possible either initially, when the check in line~\ref{rc:line:valid1} of
Algorithm~\ref{alg:RealizabilityCheck} was invalid,
or from the final deadlocked state computed in
line~\ref{rc:line:verify} of Algorithm~\ref{alg:RealizabilityCheck}
otherwise.
This last valuation is appended to the deadlocking computation at the end.

Algorithm~\ref{alg:GetDeadlockingCompAndConflictStar} first initializes $\phi$ 
with a constraint that defines the valuation
of $\vS$ for the last state in the input counterexample, when the counterexample
is not empty, or with $\top$ otherwise.
Then, it creates activation literals $L$ (line~\ref{ctrace:line:alits}) that
will be used to track the contribution of each guarantee in $G^*$ to
the unsatisfiability of
$A^* \land \varphi[\vS] \land \neg R[\vS,\nxtI] \land
\bigwedge_{g_j\in G^*} l_j \Rightarrow g_j$.
But first, the algorithm finds a valuation that maximizes the number of
satisfied guarantees in $G^*$ by solving a MaxSMT problem
consisting in the hard constraint introduced in
line~\ref{ctrace:line:constraints}, and a soft clause
for each activation literal guarding a guarantee constraint
(line~\ref{ctrace:line:soft}).
The algorithm uses the generated model $\theta$ to
fix the values for the inputs in the last step
(line~\ref{ctrace:line:inputs}).
Then, it computes a minimal set of unsatisfiable guarantees
(line~\ref{ctrace:line:core}).
In lines~\ref{ctrace:line:isEmpty2}-\ref{ctrace:line:addToCex} the algorithm
extends the input counterexample with the computed valuation.
Finally, the algorithm returns the final deadlocking computation and
the set of conflicting guarantees based on the activation literals 
included in the unsat core (line~\ref{ctrace:line:return}).

\begin{algorithm}
\caption{\textsc{GetDeadlockingCompAndConflict} ($R$, $cex$, $A^*$, $G^*$)}
\begin{algorithmic}[1]

  \If{$\proc{IsEmptyList}{cex}$} \Comment{Initial Case}

    \State $\varphi[\vS] \gets \top$
  
  \Else \Comment{Transition Case}

    \State $\sigma \gets \proc{GetLastElementOfList}{cex}$ \Comment{Map from state and input variables to values}

    \State $\varphi[\vS] \gets \bigwedge_{s_j\in\vS} s_j=\sigma(s_j)$

  \EndIf

  \State Create activation literals $L = \{l_j\mid g_j \in G^*\}$ \label{ctrace:line:alits}

  \State $\proc{SmtAssert}{A^* \land \varphi[\vS] \land \neg R[\vS,\nxtI] \land \bigwedge_{g_j\in G^*} l_j \Rightarrow g_j}$
         \label{ctrace:line:constraints}

  \State $\proc{SmtPush}$

  \For{$l_j\in L$}
    \State $\proc{SmtAssertSoft}{l_j, 1}$ \label{ctrace:line:soft}
  \EndFor

  \State $\theta \gets \proc{SmtCheckSatAndGetModel}{ }$ \label{ctrace:line:theta}

  \State $\proc{SmtPop}$

  \State $\proc{SmtAssert}{\bigwedge_{i_j\in\vI} i_j=\theta(i_j)}$
         \label{ctrace:line:inputs}

  \State $\_ \gets \proc{SmtCheckSatAssuming}{L}$ \label{ctrace:line:sat-assumming}

  \State $U \gets \proc{SmtGetMinimalUnsatCore}{ }$ \label{ctrace:line:core}

  \If{$\proc{IsEmptyList}{cex}$} \label{ctrace:line:isEmpty2}

     \State $\sigma' \gets \{ i_j \mapsto \theta(i_j) \mid i_j\in\vI\} \cup
             \{s_j \mapsto \theta(s_j) \mid s_j\in\vS\}$

  \Else

     \State $\sigma' \gets \{ i_j \mapsto \theta(i_j) \mid i_j\in\vI\} \cup
              \{s'_j \mapsto \theta(s'_j) \mid s'_j\in\vec{s'}\}$

  \EndIf

  \State $cex' \gets \proc{AddElementAtTheEnd}{\sigma', cex}$ \label{ctrace:line:addToCex}

  \State \Return $cex', \proc{MapActivationLiteralsToGuarantees}{U}$ \label{ctrace:line:return}

\end{algorithmic}
\label{alg:GetDeadlockingCompAndConflictStar}
\end{algorithm}

\section{Related Work}

The realizability check described in this report is largely based on
the synthesis procedure for infinite-state reactive systems, called \jsynvg,
presented in~\cite{Katis18}.
The only difference between both works is more practical than
theoretical.
While the original work relies on a dedicated solver to 
implement the functionality provided by the \aeval procedure~\cite{AEVAL19},
our tool only requires a generic quantifier elimination procedure
for the underlying theories supported by \kind (LIA and LRA).
These procedures are commonly available in state-of-the-art SMT solvers
like Z3~\cite{Z3} and CVC4~\cite{CVC4}.
The use of a standard solver is also the approach followed by the 
synthesis tool \gensys, recently published in~\cite{GenSys},
which was developed contemporary with our tool.
As the experimental evaluation shows later,
the use of standard solvers can improve the performance
and increase the set of solved instances on
the set of benchmarks used in the original work.

Another notable realizability check algorithm for
infinite-state specifications is the one presented in~\cite{Gacek15},
called \jsyn, which follows a k-induction approach.
Like the algorithm described in this report, it is also based on
the notion of viability explained in Section~\ref{sec:preliminaries}.
However, the algorithm suffers from soundness problems with respect to 
unrealizable results which limits its applicability.

A recent work on realizability checking of infinite-state specifications is
the compositional realizability analysis presented in~\cite{MavridouKGKPW21},
which is a preprocessing step that can be applied to assume-guarantee contracts.
It automatically partitions specifications into sets of non-interfering requirements
so that checking whether a specification is realizable reduces to checking that 
each partition is realizable.
Since this is an orthogonal technique that can improve the scalability of
the functionality provided by \kind, we will study its integration in \kind
in the future.

\section{Experimental Evaluation}

\begin{figure}[t]%
  \centering
  \includegraphics[width=0.7\textwidth]{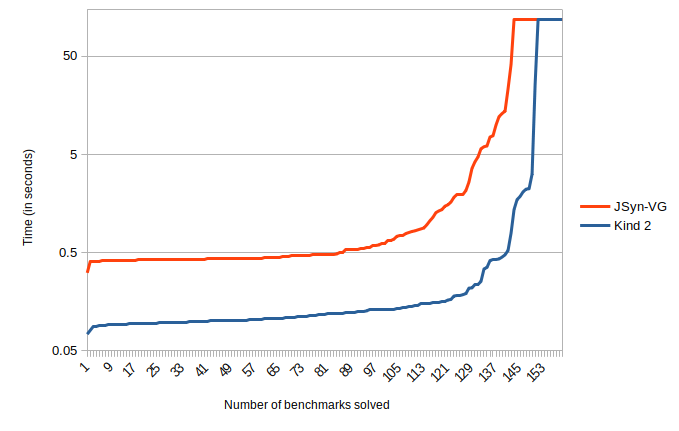}
  \caption{Comparison between \jkind (\jsynvg) and \kind}
  \label{fig:kind2-vs-jsynvg}
\end{figure}

We compared our realizability check implementation in \kind with 
the latest version of \jsynvg available within the JKind model checker
(\url{https://github.com/andrewkatis/jkind-1/releases/tag/1.8}).
We ran each tool on a Linux machine with eight 4-core Intel
i7-6700 processors and 32GB of memory using a timeout of 2 minutes.
We used the benchmarks available at
\url{https://github.com/andreaskatis/synthesis-benchmarks},
which includes the 124 contracts used in~\cite{Katis18} plus 50 more 
contracts added to the repository after the publication of the work.
Since \jkind doesn't have native support for the specification of
contracts, the benchmarks are encoded using Lustre \code{assert} statements,
and two special statements, \code{REALIZABLE} and \code{PROPERTY}.
To run \kind on the benchmarks, we encoded the problems using 
\kind built-in assume-guarantee specification language.

After running the experiments, we found that \kind rejected two of the problems
before any analysis was performed due to syntactic restrictions imposed by \kind,
and that \kind and \jkind disagreed upon the result on 13 of the problems.
The two rejected problems contained assumptions over current values of 
outputs streams, which \kind does not accept as a way of encouraging 
good practices when writing specifications. We have often found that
this kind of assumptions are not usually what the user intended to
specify and they lead to subtle flaws.
With regard to the problems were \kind and \jkind disagreed on,
they included unguarded applications of the \code{pre} operator,
which leads to undefined behavior at the initial step.
In the semantics of \jkind, each unguarded applications of
the \code{pre} operator over the \emph{same} expression is
treated as a \emph{single} undefined constant value.
In contrast, \kind's semantics treats each unguarded applications of
the \code{pre} operator as a potentially \emph{different} undefined
constant value even if it is applied to the same expression.
This leads \kind to classify as unrealizable problems that
\jkind classifies as realizable.

To make a fair comparison we decided to remove the 15 problems
mentioned above from the set of benchmarks, and carry our experimental
evaluation over the remaining 159 problems.
Moreover, the experimental evaluation only takes into account the runtime
required to determine the realizability of the contracts, and thus,
it excludes the generation of the deadlocking computation and conflict in the case
of \kind, and the synthesis of an implementation in the case of \jkind.

Figure~\ref{fig:kind2-vs-jsynvg} shows that \kind out-performances
the implementation of \jsynvg in \jkind providing an answer faster 
and in more cases.
Moreover, the set of problems solved by \kind is a strictly
larger superset of the problems solved by \jkind.
When we doubled the original timeout up to 4 minutes, \jkind was able to solve
only one more problem already solved by \kind.

In addition, we quantified the overhead of generating a deadlocking computation
and a conflict for the contracts on the benchmark set that \kind
classified as unrealizable.
Computing the additional information for the 22 contracts which \kind could
prove unrealizable took 37 seconds more, increasing the total runtime
from 87 to 124 seconds. This represents a 43\% overhead.

\bibliographystyle{splncs04}
\bibliography{main.bib}

\end{document}

%% file: macros.tex
\usepackage{fullpage}
\usepackage{graphicx}

\usepackage{cite}
\usepackage{amssymb}
\usepackage{algorithm}
\usepackage{algpseudocode}
\usepackage{xcolor}
\usepackage{listings}
\usepackage{subcaption}

\usepackage{url}
\usepackage{xspace}
\usepackage{hyperref}
\hypersetup{
 pdfborder={0 0 0},
 colorlinks=true,
 linkcolor=blue,
 urlcolor=blue,
 citecolor=blue
}

\algdef{SE}[DOWHILE]{Do}{doWhile}{\algorithmicdo}[1]{\algorithmicwhile\ #1}%
\algtext*{EndFor}
\algtext*{EndIf}

\definecolor{mymauve}{rgb}{0.58,0,0.82}

\lstdefinelanguage{lustre}{
  escapechar=|,
  keywords=[1]{
    node,function,returns,let,tel,var,const,
    assert,when,current,true,false,pre,if,
    then,else,not,and,or,contract,
    type,imported
  },
  keywords=[2]{int,real,bool,subrange},
  keywords=[3]{
    ensure, guarantee
  },
  keywords=[4]{
    require, assume
  },
  keywords=[5]{
    mode,@mode,contract,@contract,import
  },
  keywords=[6]{
    @var,@const
  },
  sensitive=true,
  morecomment=[l]{--},
  morestring=[b]",
  morestring=[b]""",
  numberstyle=\tiny\color{white!40!black},
  keywordstyle=[1]\color{blue},
  keywordstyle=[2]\color{orange}\bfseries,
  keywordstyle=[3]\color{red!60!black}\bfseries,
  keywordstyle=[4]\color{green!60!black}\bfseries,
  keywordstyle=[5]\color{orange!60!black}\bfseries,
  keywordstyle=[6]\color{blue!60!black}\bfseries,
  commentstyle=\color{white!60!black}\it\bfseries,
  stringstyle=\color{gray},
  aboveskip=1mm,
  belowskip=1mm,
  showstringspaces=false,
  numbers=left,
  numbersep=5pt,
  breaklines=true,
  breakatwhitespace=true,
  columns=fullflexible,
  keepspaces=true,
  tabsize=3,
}

\newcommand{\tinyLustreLst}[0]{
  \lstset{language=lustre, basicstyle={\footnotesize\ttfamily}}
}
\lstnewenvironment{lustreTiny}[1][0pt]{
  \tinyLustreLst
  \lstset{xleftmargin=#1} 
}{}
\lstnewenvironment{lustreTinyMath}[1][0pt]{
  \tinyLustreLst
  \lstset{mathescape, xleftmargin=#1} 
}{}

\newcommand{\code}[1]{{\texttt{#1}}}

\newcommand{\kind}{\textsc{Kind}~2\xspace}
\newcommand{\jkind}{\textsc{JKind}\xspace}
\newcommand{\aeval}{\textsc{AE-VAL}\xspace}
\newcommand{\jsyn}{\textsc{JSyn}\xspace}
\newcommand{\jsynvg}{\textsc{JSyn-vg}\xspace}
\newcommand{\gensys}{\textsc{GenSys}\xspace}

\newcommand{\define}[1]{{\emph{#1}}}

\newcommand{\false}{\mathit{false}}
\newcommand{\true}{\mathit{true}}

\newcommand{\vRS}{\hat{\vec{s}}}
\newcommand{\vX}{\vec{x}}
\newcommand{\vY}{\vec{y}}
\newcommand{\vS}{\vec{s}}
\newcommand{\vI}{\vec{i}}
\newcommand{\nxtS}{\vS'}
\newcommand{\nxtI}{\vI'}

\newcommand{\sys}{S}
\newcommand{\init}{I}
\newcommand{\trans}{T}
\newcommand{\sysTuple}{
   \langle \vS, \vI, \init[\vS,\vI], \trans[\vS, \nxtI, \nxtS] \rangle
}

\newcommand{\reachable}[1]{\textsc{Reachable}_{#1}}
\newcommand{\reachableSA}{\reachable{\sys,\ass}}
\newcommand{\viable}[1]{\textsc{Viable}_{#1}}
\newcommand{\viableC}{\viable{\agContract}}

\newcommand{\agContract}{C}
\newcommand{\ass}{A}
\newcommand{\env}{E}
\newcommand{\gua}{G}
\newcommand{\assI}{\ass_I}
\newcommand{\assT}{\ass_T}
\newcommand{\envI}{\env_I}
\newcommand{\envT}{\env_T}
\newcommand{\guaI}{\gua_I}
\newcommand{\guaT}{\gua_T}
\newcommand{\agTuple}{
   \langle \ass, \gua \rangle
}
\newcommand{\assTP}{
  \assT[\vS, \nxtI]
}
\newcommand{\assTuple}{
   \langle \vI, \vS, \assI[\vI], \assTP \rangle
}
\newcommand{\envTuple}{
   \langle \vI, \vS, \envI[\vI], \envT[\vS, \nxtI] \rangle
}
\newcommand{\guaTuple}{
   \langle \vS, \vI, \guaI[\vS,\vI], \guaT[\vS, \nxtI, \nxtS] \rangle
}

\newcommand{\emptyTuple}{
  \langle \vS, \vI, \top, \top \rangle
}

\newcommand{\validRegionP}{
  validRegion[\vS,\nxtI]
}

\newcommand{\pair}[2]{
  \langle #1, #2 \rangle
}

\newcommand{\proc}[2]{\ensuremath{\mathsf{#1}}(#2)}
\newcommand{\verify}{\ensuremath{\mathsf{Verify}}\xspace}

\newcommand{\mvar}[1]{\mathit{#1}}

\newtheorem{mydef}{Definition}
